\documentclass[twocolumn,superscriptaddress,pra]{revtex4}
\usepackage[koi8-r]{inputenc}
\usepackage[english,russian]{babel}
\usepackage{amsmath}
\usepackage{amssymb}
\usepackage{epsfig}

\begin{document}
\title{``Точные'' решения для циркулярно поляризованных Керровских солитонов}

\author{В.\,П.\,Рубан}
\email{ruban@itp.ac.ru}
\affiliation{Институт теоретической физики им.~Л.\,Д.\,Ландау РАН, 142432, 
Черноголовка, Россия}

\author{Р.\,В.\,Рубан}
\affiliation{Физический факультет, МГУ им.~М.\,В.\,Ломоносова, 119991, 
Москва, Россия}

\date{\today}

\begin{abstract}
Для нелинейного векторного уравнения типа ротор-ротор, описывающего
монохроматическую световую волну в Керровской среде, предлагается точная 
редукция, приводящая к системе четырех обыкновенных дифференциальных
уравнений первого порядка для функций от поперечной радиальной координаты.
Численные решения этой системы с подходящими краевыми условиями дают
исчерпывающую информацию о внутреннем строении сильно нелинейного 
стационарного оптического пучка, имеющего в основном определенную круговую
поляризацию, но с небольшой ``примесью'' двукратно завихренной противоположной
поляризации и однократно завихренной продольной компоненты электрического поля.

\end{abstract}

\maketitle

{\bf Введение.}
В нелинейной оптике важную роль играет векторное уравнение 
\begin{equation}
\mbox{rot}\, \mbox{rot}\, {\bf E}=\frac{\omega^2}{c^2}[\varepsilon{\bf E}
+\alpha |{\bf E}|^2{\bf E}+\beta({\bf E}\cdot{\bf E}){\bf E}^*],
\label{curl_curl}
\end{equation}
которое описывает строго монохроматическую световую волну с частотой $\omega$ 
в прозрачной оптической среде с Керровской нелинейностью \cite{LL8}. Здесь 
${\bf E}(x,y,z)$ --- комплексная амплитуда главной гармоники электрического поля, 
$\varepsilon(x,y,z)$ --- диэлектрическая проницаемость для волн малой амплитуды
на заданной частоте, $\alpha(x,y,z)$ и $\beta(x,y,z)$ --- коэффициенты нелинейности.
Для пространственно однородной среды $\varepsilon$, $\alpha$ и $\beta$
--- постоянные параметры (причем в пределе мгновенного нелинейного отклика
Керровские коэффициенты связаны соотношением $\beta=\alpha/2$). В случае 
положительных $\alpha$ и $\beta$ (фокусирующая среда), среди решений уравнения
(\ref{curl_curl}) имеются, в частности, т. н. ``световые иглы'' --- световые пучки 
шириной порядка длины волны $\lambda_0=2\pi/k_0$, где 
$k_0=\sqrt{\varepsilon}\omega/c$ \cite{opt_n_1,opt_n_2,opt_n_3,opt_n_4}. 
Соответственно, в дефокусирующей среде (отрицательные $\alpha$ и $\beta$) возможны
оптические вихри \cite{opt-vort-1,opt-vort-2,OptVort,VortNonlFields,opt-vort-3}, 
а также доменные стенки между областями с противоположной круговой поляризацией света
\cite{PDW1,PDW2,PDW3,PDW4,PDW5,dom_wall-1,dom_wall-2}. 
Все эти объекты представляют большой теоретический и практический интерес.

Как правило, аналитические исследования уравнения (\ref{curl_curl}) проводятся
приближенными методами, имея в качестве ``отправной точки'' систему связанных
нелинейных уравнений Шредингера (НУШ). В главном приближении амплитуды 
$A_\pm(x,y,z)$ левой и правой круговых поляризаций в выражении
\begin{equation}
{\bf E}\approx \big[({\bf e}_x+i{\bf e}_y) A_+ 
+ ({\bf e}_x-i{\bf e}_y) A_- \big]\exp(ik_0 z)/\sqrt{2} 
\end{equation}
предполагаются медленными функциями, при полном пренебрежении малой дивергенцией
электрического поля и его продольной компонентой. В этих предположениях
получается пара НУШ \cite{BZ1970}
\begin{equation}
-2i k_0 \partial_z A_\pm=\Delta_\perp A_\pm 
+\frac{\alpha k_0^2}{\varepsilon}\Big[|A_\pm|^2 + g|A_\mp|^2\Big]A_\pm,
\label{A_pm_eqs}
\end{equation}
где параметр перекрестной фазовой модуляции $g=1+2\beta/\alpha$. Затем 
рассматриваются поправки по малому отношению длины волны к характерной ширине
пучка, которые учитывают $\mbox{div}\,{\bf E}\neq 0$ и $E_\parallel \neq 0$
\cite{MNLSE-1,MNLSE-2,MNLSE-3,beam1,beam2,beam3,beam4,beam5}. Для большинства
практических целей этого оказывается вполне достаточно, поскольку уменьшение
ширины пучка до величин порядка длины волны неизбежно подразумевает увеличение 
амплитуды $E\equiv|{\bf E}|$ поля до таких значений, когда в правой части
уравнения (\ref{curl_curl}) нелинейные слагаемые оказываются сравнимыми с линейным
членом, то есть $\alpha E^2\sim \varepsilon$. Но в обычных веществах с 
$\varepsilon\sim 1$ такое соотношение никогда не выполняется при умеренных
интенсивностях света, а в сильных полях уже сама модель (\ref{curl_curl}) 
становится непригодной, так как включаются неучтенные ею физические процессы
\cite{filamentation}.
Однако имеются теоретические оценки, что в искусственно созданных композитных
материалах могут достигаться достаточно малые значения $\varepsilon$ (а также
параметр $\gamma=\beta/\alpha$ может задаваться в широких пределах)
\cite{composite-1,composite-2,composite-3}. В таком случае анализ точных солитонных
решений уравнения (\ref{curl_curl}) с большими амплитудами не только интересен с 
научной точки зрения, но и практически важен. Этой проблеме и посвящена данная работа.

{\bf Редукция.}
Как известно, при изучении нелинейных систем в частных производных желанной целью 
и большой удачей для исследователей всегда оказывается нахождение точных редукций 
--- самосогласованных подстановок, которые сводят задачу к системе обыкновенных
дифференциальных уравнений (ОДУ). Например, одна из немногих известных редукций
уравнения (\ref{curl_curl}) --- осесимметричные решения в виде вихря с азимутальной
поляризацией в дефокусирующей среде \cite{azimuth-1,azimuth-2}. 
В фокусирующем случае основной интерес представляют пространственные солитоны 
--- локализованные по обоим поперечным направлениям устойчивые нелинейные волновые
пакеты. В рамках связанных НУШ такие двумерные объекты, как известно, не существуют,
поскольку волновой пакет либо расплывается за счет дифракции, либо коллапсирует 
за счет нелинейности (см. \cite{Berge1998,ZK2012UFN} и ссылки там). Но с учетом
нелинейно-дифракционных поправок солитоны возможны, причем при произвольном
соотношении между левой и правой круговыми поляризациями в световом пучке, включая
крайние случаи --- линейную и круговую поляризации. Надо сказать, что 
с точки зрения поиска точных редукций линейная и эллиптическая поляризации
не выглядят перспективными, поскольку там профиль пучка заведомо не обладает
вращательной симметрией в поперечной плоскости. Совсем другое дело --- пучки 
с круговой поляризацией. В данной работе предлагается подстановка, которая сводит
задачу о поперечном строении такого пучка к системе четырех нелинейных ОДУ 
первого порядка для неизвестных функций радиальной координаты $r=\sqrt{x^2+y^2}$.
Более того, число участвующих в решении свободных параметров оказывается достаточным
для того, чтобы обеспечить регулярность всех функций при $r\to 0$ и их нулевую
асимптотику при $r\to \infty$. Таким образом, нами здесь получена детальная информация
о пространственных Керровских солитонах не только малой, но и конечной амплитуды.

Подстановка, о которой идет речь, применима не только непосредственно к уравнению
(\ref{curl_curl}), но и к его обобщениям вида
\begin{eqnarray}
&&\mbox{rot}\, \mbox{rot}\, {\bf E}=
N(r, E^2, |{\bf E}\cdot{\bf E}|^2){\bf E}\nonumber\\
&&\qquad+\tilde M(r, E^2, |({\bf E}\cdot{\bf E})|^2) ({\bf E}\cdot{\bf E}){\bf E}^*,
\label{curl_curl_NM}
\end{eqnarray}
где $N$ и $\tilde M$ --- некоторые заданные вещественные функции указанных
скалярных аргументов. Заметим попутно, что в случае 
$$
N{\bf E}+\tilde M({\bf E}\cdot{\bf E}){\bf E}^*=\partial \Pi/\partial{\bf E}^*
$$
данное уравнение относится к вариационному типу. Для обычной Керровской нелинейности
$N=k_0^2[1 +(\alpha/\varepsilon) E^2]$ и $\tilde M=k_0^2(\beta/\varepsilon)=$ const, 
так что 
$$
\Pi=k_0^2[E^2+(\alpha/\varepsilon) E^4/2+(\beta/\varepsilon)|({\bf E}\cdot{\bf E})|^2/2].
$$
Можно также рассматривать и среды с насыщающейся нелинейностью
\cite{saturat-1,saturat-2,saturat-3}.

Мы будем искать точные решения уравнения (\ref{curl_curl_NM}) в виде 
${\bf E}=\hat{\bf E}(r,\varphi)e^{i\kappa z}$, где $\varphi$ --- полярный угол,  
$\kappa$--- константа распространения. Поперечный профиль пучка попробуем задать
следующей подстановкой:
\begin{eqnarray}
\hat{\bf E}&=&{\bf e}_x[A_+(r)+A_-(r)e^{2i\varphi}]\nonumber\\
&+&i{\bf e}_y[A_+(r)-A_-(r)e^{2i\varphi}]-i{\bf e}_z B(r)e^{i\varphi}.
\label{ansatz}
\end{eqnarray}
Здесь $A_+(r)$ --- лево-поляризованная компонента волны, которую мы предполагаем
доминируюшей, $A_-(r)e^{2i\varphi}$ --- относительно малая ``примесь'' 
право-поляризованной компоненты, содержащая двукратный вихрь, 
$-i B(r)e^{i\varphi}$ --- продольное поле, содержащее однократный вихрь.
Легко видеть, что $E^2=2|A_+|^2+2|A_-|^2+|B|^2$,
$({\bf E}\cdot{\bf E})=(4A_+ A_- -B^2)e^{2i\varphi}$.

После соответствующих вычислений мы находим, что $x$-компонента уравнения дает
условие вида $W_0(r)+W_2(r)e^{2i\varphi}=0$, с некоторыми дифференциальными 
выражениями $W_0$ и $W_2$, тогда как для $y$-компоненты получается совместное с
предыдущим уравнение $i[W_0(r)-W_2(r)e^{2i\varphi}]=0$. Для продольной компоненты
мы имеем $iW_1(r)e^{i\varphi}=0$. Таким образом, число независимых уравнений 
($W_0=0$, $W_1=0$ и $W_2=0$) совпадает с числом неизвестных функций.

По техническим соображениям удобно ввести линейные комбинации $(A_+ +A_-)=F$ и
$(A_+ -A_-)=G$. При этом $E^2=|F|^2+|G|^2+|B|^2$, 
$({\bf E}\cdot{\bf E})=(F^2-G^2-B^2)e^{2i\varphi}$.
Тогда получается система ОДУ следующего вида:
\begin{equation}
\kappa^2 F+\frac{(F-G)}{r^2}-\frac{G'}{r}+\kappa B'=NF+M F^*,
\label{F_eq}
\end{equation}
\begin{equation}
\kappa^2 G -G''+\Big[\frac{(F-G)}{r}\Big]'+\kappa \frac{B}{r}
=NG-M G^*,
\label{G_eq}
\end{equation}
\begin{equation}
-\frac{(rB')'}{r}+\frac{B}{r^2}-\kappa\Big[F'+\frac{(F-G)}{r}\Big]=NB-M B^*,
\label{B_eq}
\end{equation}
где для краткости введено обозначение 
$$
M=(F^2-G^2-B^2)\tilde M.
$$

Дальнейшее упрощение полученной системы возможно в силу равенства 
$\mbox{div}\,[N{\bf E}+\tilde M({\bf E}\cdot{\bf E}){\bf E}^*]=0$,
которое прямо следует из уравнения (\ref{curl_curl_NM}). Составив комбинацию
$[r\cdot$Ур.(\ref{F_eq})$]'-$Ур.(\ref{G_eq})$+\kappa r\cdot$Ур.(\ref{B_eq}),
получим ОДУ первого порядка
\begin{eqnarray}
&&(NF+M F^*)'+\frac{1}{r}[N(F-G)+M(F^*+G^*)]\nonumber\\
&&\qquad+\kappa(NB-M B^*)=0,
\label{div_D}
\end{eqnarray}
которое заменяет собой уравнение (\ref{B_eq}).

\begin{figure}
\begin{center} 
\epsfig{file=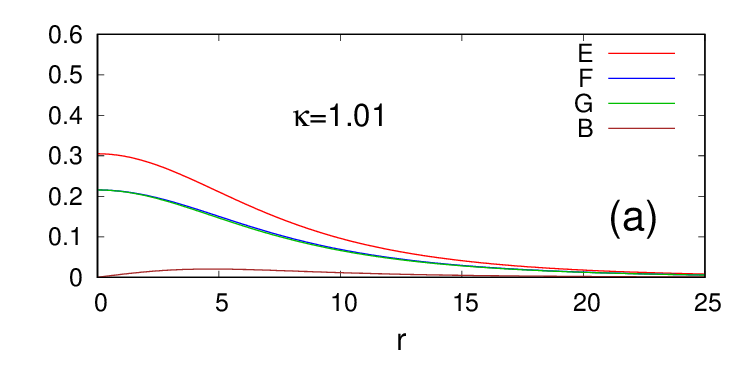, width=86mm}\\
\epsfig{file=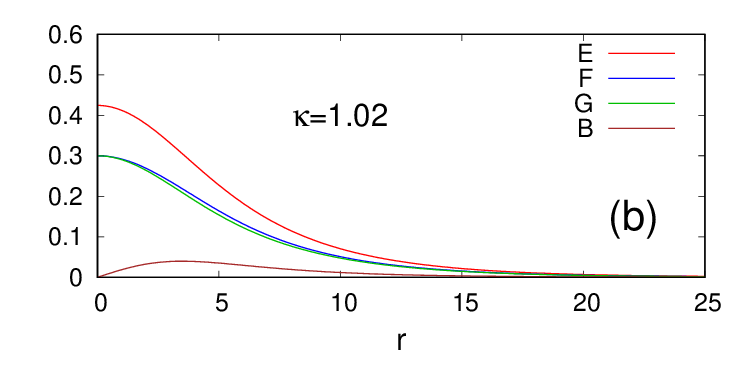, width=86mm}\\
\epsfig{file=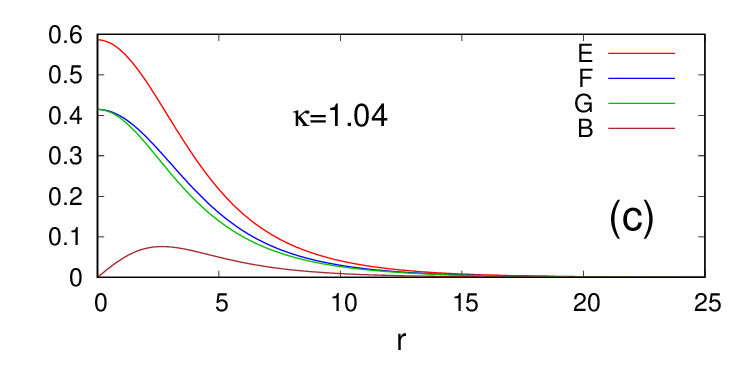, width=86mm}\\
\epsfig{file=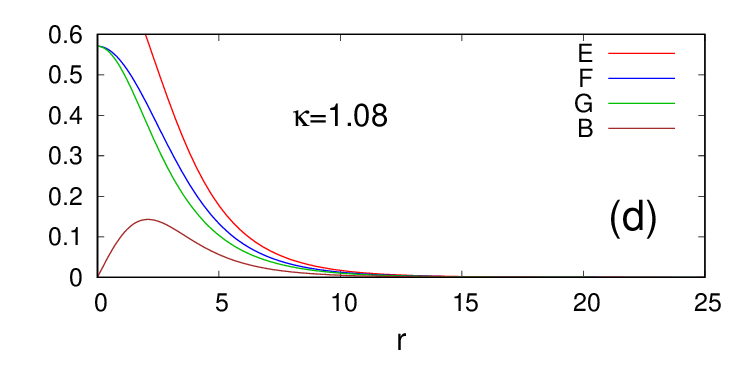, width=86mm}
\end{center}
\caption{Примеры профилей Керровских солитонов. Соответствующие значения параметров:
a) $\kappa = 1.01$, $A_0 = 0.215682$,  $b_1 = 0.007626$;
b) $\kappa = 1.02$, $A_0 = 0.3000665$, $b_1 = 0.0198068$;
c) $\kappa = 1.04$, $A_0 = 0.4145696$, $b_1 = 0.0492948$;
d) $\kappa = 1.08$, $A_0 = 0.57056007$, $b_1 = 0.11817944$.
}
\label{solitons} 
\end{figure}

{\bf Способ численного решения и примеры.}
Далее мы будем предполагать функции $F$, $G$ и $B$ действительными,
что допускается системой.

В уравнении (\ref{F_eq}) производная $B'$ легко выражается через $\{F,G,G',B\}$.
Затем из уравнений (\ref{div_D}) и (\ref{F_eq}) можно алгебраически выразить 
$F'$ также через указанную четверку функций. В итоге мы имеем систему ОДУ 
стандартного типа в четырехмерном фазовом пространстве, которая легко решается
численными методами с высокой точностью. Решать приходится, однако, не задачу Коши
с начальными условиями, а краевую задачу. Если мы рассматриваем решения на отрезке
между $r_1>0$ и $r_2>r_1$, то естественные краевые условия, которые соответствуют
идеально проводящим стенкам волновода, состоят в требованиях $G(r_1)=0$, $G(r_2)=0$,
$B(r_1)=0$, $B(r_2)=0$, и в свободных краевых условиях для $F(r_1)$ и $F(r_2)$.
При этом касательная компонента электрического поля обращается в ноль на 
цилиндрических стенках волновода. Если же рассматриваются солитоны в безграничном
пространстве, то на бесконечности следует поствить нулевые асимптотические условия,
а при $r\to 0$ все наши функции обязаны быть регулярными. 

Чтобы использовать численные методы интегрирования полученных ОДУ
(например, как в нашем случае, метод Рунге-Кутта 4-го порядка аппроксимации), 
необходимо хотя бы слегка ``отойти'' от особой точки $r=0$. Для этого надо знать
несколько первых коэффициентов в разложениях неизвестных функций по степеням $r$,
\begin{eqnarray}
F&=&A_0+\frac{f_2}{2}r^2+\cdots,\\
G&=&A_0+\frac{g_2}{2}r^2+\cdots,\\
B&=&b_1 r +\cdots.
\label{expansion}
\end{eqnarray}
Весьма примечательно, что оба уравнения (\ref{F_eq}) и (\ref{G_eq}) дают одно и то же
требование на коэффициенты $A_0$, $f_2$, $g_2$ и $b_1$:
\begin{equation}
\frac{1}{2}f_2-\frac{3}{2}g_2+\kappa b_1=[N(0,2A_0^2,0)-\kappa^2]A_0.
\end{equation}
Еще одна связь между коэффициентами разложения получается из уравнения (\ref{div_D}).
Ввиду громоздкости мы ее здесь не выписываем явно.

Таким образом, на четыре коэффициента имеется всего два алгебраических уравнения.
Удобно взять $A_0$ и $b_1$ в качестве свободных параметров, а через них выразить 
$f_2$ и $g_2$. Подбирая затем нужным образом $A_0$ и $b_1$, можно добиться выполнения
нулевых асимптотических условий на бесконечности. При этом требуется весьма 
``тонкая'' совместная настройка значений $A_0$ и $b_1$, для чего приходится 
проводить численную процедуру типа ``стрельбы''. Несколько численных примеров
профилей солитонов приведены на Рис.\ref{solitons} для обезразмеренного уравнения
(\ref{curl_curl}), что формально соответствует выбору $N=1+E^2$, $\tilde M=\gamma=0.5$
(при этом $k_0=1$, $\lambda_0=2\pi$). Из этого рисунка видно, что примесь 
противоположной поляризации $A_-=(F-G)/2$ становится заметной лишь для весьма узких
солитонов. Продольная компонента электрического поля также всегда относительно мала.

{\bf Заключение.} Точное приведение векторной задачи о циркулярно поляризованном,
самоподдерживающемся нелинейном световом пучке к системе нескольких ОДУ, выполненное
в данной работе, представляется значительным теоретическим результатом. 
Эта же система ОДУ, но с дефокусирующей нелинейностью, должна дать и решение задачи
о двукратном вихре в правой компоненте, ядро которого содержит левую компоненту. 
Кроме того, есть смысл рассмотреть в будущем подстановки типа (\ref{ansatz}), 
но с дополнительным общим множителем $e^{im\varphi}$, при целочисленных $m$. 
Например, известные решения с азимутальной поляризацией \cite{azimuth-1,azimuth-2}
должны возникать при $m=-1$. 

Представляется также важной задачей провести исследования устойчивости решений с 
учетом вариационного принципа для поля $\hat{\bf E}(x,y)$, которое, как легко видеть,
минимизирует интеграл
\begin{equation}
\int\Big[|\partial_x\hat E_y-\partial_y\hat E_x|^2
+|\kappa \hat{\bf E}+i\nabla_\perp \hat E_z|^2
-\Pi(\hat{\bf E},\hat{\bf E}^*)\Big] dx dy.
\end{equation}
Возможно, на этом пути удастся добиться более высокого уровня в качественном 
понимании свойств мелкомасштабных световых структур.

\end{document}